# Cross Observatory Coordination with `tilepy`: A Novel Tool for Observations of Multimessenger Transient Events

Monica Seglar-Arroyo[1], Halim Ashkar[2], Mathieu de Bony de Lavergne[3], and Fabian Schüssler[3]
[1] IFAE, The Barcelona Institute of Science and Technology, Campus UAB, 08193 Bellaterra, Barcelona, Spain; astro.tilepy@gmail.com
[2] Laboratoire Leprince-Ringuet, École Polytechnique, CNRS, Institut Polytechnique de Paris, F-91128 Palaiseau, France
[3] IRFU, CEA, Université Paris-Saclay, F-91191 Gif-sur-Yvette, France

## Abstract

Time-domain astrophysics has leaped forward with the direct discovery of gravitational waves and the emergence of new generation instruments for multimessenger studies. The capacity of the multimessenger multiwavelength community to effectively pursue follow-up observations is hindered by the suboptimal localization of numerous transient events and the escalating volume of alerts. Thus, we have developed an effective tool to overcome the observational and technical hurdles inherent in the emerging field of multimessenger astrophysics. We present `tilepy`, a Python package for the automatic scheduling of follow-up observations of poorly localized transient events. It is ideally suited to tackle the challenge of complex follow-up in mid- and small-field-of-view telescope campaigns, with or without human intervention. We demonstrate the capabilities of `tilepy` in the realm of multiobservatory, multiwavelength campaigns, to cover the localization uncertainty region of various events ultimately aiming at pinpointing the source of the multimessenger emission. The `tilepy` code is publicly available on GitHub and is sufficiently flexible to be employed either automatically or in a customized manner, tailored to collaboration and individual requirements. `tilepy` is also accessible via a public API and through the Astro-COLIBRI platform.

*Unified Astronomy Thesaurus concepts:* Observational astronomy (1145); Astronomical methods (1043); Transient detection (1957); Gamma-ray bursts (629); Gravitational wave sources (677)

## 1. Introduction and Science Drivers

Over the last decade, a new era of transient multimessenger astrophysics has been established with the discovery of a gamma-ray burst (GRB) counterpart to the gravitational wave (GW) GW170817 event (Abbott et al. 2017) and the hint of a dual detection of neutrinos and gamma-ray flares from the TXS 0506 + 056 (IceCube Collaboration et al. 2018) blazar. The ability to detect an astrophysical source through diverse channels enhances our capacity to gain a comprehensive understanding of the underlying phenomena. This approach can unveil crucial details including acceleration processes, energetics, environmental conditions, dynamics, and the masses and orientations of the sources. This realization prompts the community to conduct a rapidly increasing number of follow-up observations in pursuit of additional multimessenger detections.

These attempts face two main challenges: the necessity to react swiftly to catch the often rapidly fading emission associated with transient astrophysical phenomena and the need to cover large regions of the sky due to the fact that a large fraction of transients are only poorly localized during initial observations. Occasionally, the localization uncertainty may extend across hundreds or even thousands of square degrees in the sky. Since most follow-up instruments have smaller fields of view than the localization regions of these poorly localized events, dedicated follow-up strategies are required. These strategies aim to efficiently cover the large localization regions, optimize the available telescope time, and point the observatories to the regions that are most likely to contain the origin of the phenomenon. The strategies need to take into account the individual characteristics of each follow-up instrument, including its duty cycle, field of view (FoV), and mode of operation.

We created `tilepy`, a Python library to tackle these challenges and optimize available follow-up resources in searches for multimessenger emissions from poorly localized (transient) phenomena. `tilepy` is adapted for instruments with small (FoV < 1°), medium (1° < FoV < 10°), or large fields of views (FoV > 10°). `tilepy` was initially developed to search for electromagnetic counterparts from GW events detected during the Observing Run O2 (2016 November–2017 August) of the LIGO–Virgo Collaboration (Seglar-Arroyo & Schüssler 2017; Ashkar et al. 2021). In its current version, the usage has been expanded to any poorly localized event whose uncertainty region is provided by a `HEALPix` (Górski et al. 2005; Zonca et al. 2019) map. `tilepy` provides ground-based multiobservatory, multitelescope, and multiwavelength scheduling of follow-up observations and can be used to schedule low-latency observations for many science cases as described in the following. `tilepy` will automatically derive an optimal follow-up observation plan for a given time, prioritizing the most probable regions to host the astrophysical event while taking telescope observability and visibility constraints into account.

### 1.1. Gravitational Waves

The main source of GWs detectable by the current generation of interferometers, observing in the Hz to kHz frequency band, is compact binary coalescences (CBCs), where each compact object can be either a black hole or a neutron star. Upon the detection of a GW event, the LIGO–Virgo–KAGRA Collaboration distributes open public alerts[4] containing the GW localization map and various parameters describing the event.

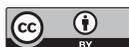

---
[4] https://emfollow.docs.ligo.org/userguide/

These maps are provided in HEALPix format (Górski et al. 2005; Zonca et al. 2019). CBC searches, which are based on match filtering of the waveform to a waveform template bank, provide 3D posterior distribution on the localization of the source. These sky maps are created by the rapid CBC sky-localization algorithm, BAYESTAR (Singer & Price 2016). Parameter estimation pipelines such as Bilby and RapidPE-RIFT are used to provide CBC sky localization within minute timescales. Burst events are searched with coherent WaveBurst, which is based on the search for coherent power excess among interferometers. These pipelines provide localization maps that only contain information on the sky-localization probability distribution of the event and no information on the luminosity distance of the source. The burst refined sky localization is in charge of LALInference Burst and Bayeswave. See https://emfollow.docs.ligo.org/userguide/ for further details on these pipelines.

tilepy has been used in GW counterpart searches by gamma-ray observatories since the LIGO–Virgo Observing Run O2. Among the various examples of these first searches, including the first three detector interferometer detection GW170817, the multimessenger campaign on GW170817 with H.E.S.S. stands out. For this event, H.E.S.S. was the first ground-based instrument to observe the true location of the binary neutron star (BNS) merger, source of the sGRB GRB170817A (Abdalla et al. 2017), before the detection of the optical counterpart in the form of a kilonova (Cowperthwaite et al. 2017). Using its galaxy-distribution informed algorithm, tilepy scheduled three observations during darkness, the first of which included the host galaxy of the BNS-sGRB (Hjorth et al. 2017). tilepy was used also for the H.E.S.S. follow-up of various binary black hole mergers during O3 (Abdalla et al. 2021; Ashkar et al. 2021). It is currently used by the CTAO-Large-Sized Telescope (LST; Carosi et al. 2021) and H.E.S.S. collaborations (Hoischen et al. 2022) for the follow-up of GW events detected during the Observing Run O4.

### 1.2. Gamma-Ray Bursts

GRB detection techniques can vary from one instrument to another. The best localization of GRBs from the current detection instrument are those provided by Swift's Burst Alert Telescope instrument (Krimm et al. 2013), which uses the coded mask technique and provides GRB localizations on the arcminute scale (Barthelmy et al. 2005). Yet, other detectors do not have the capability of localizing a signal to the subdegree precision, as is the case of Fermi's Gamma Ray Burst Monitor (GBM; Meegan et al. 2009). Following the technique pioneered by KONUS and based on the BATSE algorithm, the Fermi-GBM source localization method is based on the relative differences of rates of scintillation among detectors, which are oriented in different directions (Connaughton et al. 2015). The achieved localization uncertainty ranges from tens of square degrees to 1000 $\deg^2$ in the sky when including statistical and systematic uncertainties (Goldstein et al. 2020). The localization is provided in HEALPix format in the final notices of Fermi-GBM GRB alerts. Many other instruments have similar techniques that yield relatively poor localizations: GECAM (Zhao et al. 2023), the future SVOM-GBM (Götz et al. 2009), BurstCube (Racusin et al. 2017), and other cube satellites. The conventional approach to this challenge involved pursuing the coordinates of the best-fit position specified in the alert notification, even though the likelihood of the source being at those coordinates is often low. tilepy offers a solution by enhancing the chances of covering the plausible region from which the GRB signal originated, accounting for both statistical and systematic errors. An automatic tiling scheme using tilepy has been implemented to follow poorly localized GRBs, in the H.E.S.S. (Hoischen et al. 2022) and LST (Carosi et al. 2021) collaborations, mostly focusing on the sky map provided in Fermi-GBM final notices.

Another source of localization for GRBs is the Interplanetary Gamma-Ray Burst Timing Network (IPN). This system relies on the detection of the GRB by several instruments on well-separated spacecraft (including probes in orbit around other planets like Mars) that allow triangulating the direction of the signal (Hurley et al. 2011). HEALPix maps with the localizations performed by the network are produced and can be used in tilepy to generate an observation schedule.

### 1.3. Neutrino Candidates

High-energy astrophysical sources such as tidal disruption events, active galactic nuclei (AGNs), and stellar explosions are candidates for neutrino emission. Neutrino telescopes, such as IceCube and KM3NET, detect neutrinos in the TeV–PeV energy range via neutral and charged current interactions in the surrounding medium. Muons resulting from charged current interactions give rise to track-like event signatures that can be reasonably well reconstructed. The resulting uncertainties are typically less than 1° radius. Neutral current interactions on the other hand produce cascade (or shower)-like event signatures that result in significantly larger uncertainties on the neutrino direction (up to several tens of degrees). Both event types suffer from systematic uncertainties mainly due to the detailed characterization of the instrumented volume. The IceCube collaboration has been announcing detections of high-energy events publicly since 2016 (Aartsen et al. 2017). These alerts are crucial for near-real-time searches for electromagnetic counterparts across all wavelengths. Given the sizeable uncertainties on the reconstructed sky position of the neutrino origin, tiling strategies are often necessary to cover the region efficiently. IceCube is already publishing HEALPix localization sky maps for cascade-like events. tilepy is able to handle these sky maps and successfully obtain an optimized observation plan.

### 1.4. Extended Sources in the Sky

So far, tilepy has predominantly been used to handle complex observation scheduling in time-domain astrophysics. Nevertheless, its scope can effortlessly be expanded to scheduling observations of large sky regions, as might be the case in source morphology studies of extended sources or opportunistic scans of a priori empty sky regions. The only technical requirement is that the sky map of the region to scan is given or can be translated into the commonly used HEALPix format. tilepy will manage the evolution of the accessible sky throughout the accessible observation window and provide a priority-ordered list of observations to perform.

## 2. Functionality

### 2.1. Observability and Visibility Consideration

Observability considerations encompass the essential requirements that must be taken into account for the majority of astronomical observations, ensuring the effective operation of a specific instrument design. `tilepy` supports most types of observability considerations. They are fully customizable by the user. Darkness requirements are determined by the position of the Sun and the Moon in the sky of the telescope location. Darktime observation requires that the Sun and the Moon be at a minimum angle below the horizon. *Darktime observations* are for example favored by Imaging Atmospheric Cherenkov Telescopes (IACTs), as these detect the faint blue Cherenkov light produced by charged particles in the showers initiated by the gamma rays. In addition, many observatories (e.g., optical instruments) can conduct *Moon-time observations* that only require that the Sun be below the horizon but allow for varying conditions of Moon presence and illumination. These conditions are typically dependent on the Moon's altitude in the sky, the Moon phase, and its separation from the target source. For the latter, and in the case of poorly localized events, regions too close to the Moon, where significant moonlight background exists, can be masked and excluded from consideration for observations. Then, daytime observations do not require any conditions on the position of the Sun and the Moon. Radio telescopes that do not have any darkness requirements can operate in daytime observing mode.

Visibility or accessible sky considerations are defined by the portion of the sky that is accessible to a telescope at a certain time. It is determined by the minimum altitude angle (or maximum zenith angle) of a celestial source, indicating the point in the sky at which the telescope can effectively operate and acquire good-quality data. As an example, the visibility and observability consideration for seven different sites around the Earth, for the GW, GW170817, through the night of 2017 August 17, is provided in Figure 1.

### 2.2. Test Grid

`tilepy` searches for the highest-probability region to host the transient event to observe. Depending on the science case, the sky map can reach very high resolution, showing a highly accurate probability density distribution as for GW sky maps. Multiresolution sky maps with adaptive-mesh pixelization are being provided to take full advantage of these features without increasing the resulting file size (Martinez-Castellanos et al. 2022). The identification of the precise highest-probability sky regions is done by comparison among all the possible observations, for which a grid is used, i.e., the sky is binned. Depending on the size of the telescope FoV, we would either use the pixels themselves (or the host galaxies themselves, in galaxy-targeted approaches of small-FoV telescopes) after a proper rebinning of the sky map, or one could directly use a grid. Whenever a grid approach is used, the probabilities within the FoV will be integrated and associated with the specific grid coordinates. For each point on the grid, the probability will be assessed with the methods introduced in Section 2.3.

Three methods have been developed to create the grid. The first option degrades the resolution of the GW localization map and uses the center of the pixels in this low-resolution map. The second option allows the user to use their own grid that suits the geometry of their telescope. These options have the advantage that their resolution can be adapted to the size of the telescope's FoV or the accuracy required, which will allow for shortening the total computation time. Note that very low resolutions, although decreasing the computation time, will introduce unwanted effects and systematic uncertainties, and may not permit reaching the expected accuracy. As an example, an adequate parallel grid for a $\sim 2°$ FoV telescope corresponds to $\texttt{NSIDE} = \sim 256$ (resolution of $\sim 0°\!.22$ per pixel). The resolution of the grid allows an interplay between the accuracy of probability coverage and speed as discussed in Ashkar et al. (2021). A third option for the grid determination, feasible whenever distance information is available for the event, is to use the most probable galaxies inside the localization region considered as nodes of the grid. This option yields an inhomogeneous concentration of test points biased toward galaxy clusters and groups. We note that, for the examples we provide in this paper, we use the GLADE+ galaxy catalog (Dálya et al. 2022). Galaxies are selected based on specific scientific criteria, typically limited to certain distance layers to streamline the data set while maintaining relevance to the research objectives. In this galaxy-informed case, there are two possibilities for the grid used to speed up the computation: a grid center in the galaxy itself, and the use of a low-resolution map whose pixels are used to construct the grid. In both methods, the probability per galaxy is integrated in an area of the size of the telescope's FoV, centered around the grid nodes. A study of the performance of these methods used to build the grid showed comparable results (Ashkar et al. 2021).

### 2.3. Optimizing Sky Region Prioritization through Probability Analysis

The main information used to observe poorly localized events is the sky region defined by the probability distribution of the localization itself. The simplest way to handle these potentially large regions is to use the 2D probability as a proxy for your search. In this way, the most probable pixel or region is identified and selected to be part of the list of observations by the pointing telescope. In `tilepy`, the list of individual observations is built up by an iterative procedure that masks previously identified regions, and searches for the next one that fulfills the maximum probability in the pixel (or in the region) criteria. As explained in Section 2.2, we adapt the use of pixels or galaxies to the science case and the telescope FoV, i.e., the grid size is adapted to be comparable to the size of the telescope FoV.

More complex considerations are taken whenever a 3D probability density, corresponding to the distance to the source, can be used. In that case, we can correlate it with additional information, which is in most use-cases the distribution of galaxies within the plausible 3D-region, as done in GW counterpart searches (Singer et al. 2016). In this way, the probability of a galaxy being the host of the cataclysmic event is obtained. In `tilepy`, the galaxy catalog is defined by the user as an input argument. The GLADE+ catalog (Dálya et al. 2022) is generally used for most electromagnetic counterpart searches.

1. A galaxy-targeted search (small-FoV observatories) is a selection based on the largest galaxy.
2. A galaxy clustering search (mid/large-FoV observatories) is an integration of the convoluted probability of all the galaxies within the FoV of the telescope.

Furthermore, an additional astrophysically motivated weighting can be considered for prioritization of some classes of galaxies over others. `tilepy` provides the option to incorporate an estimate of the individual galaxy stellar masses when calculating the probability of each galaxy hosting the transient event. The stellar mass weight is assessed following Ducoin et al. (2020). Additional parameters, such as the neutron star merger rate, the metallicity, the star formation rate, etc., will be considered in the future for the probability calculation.

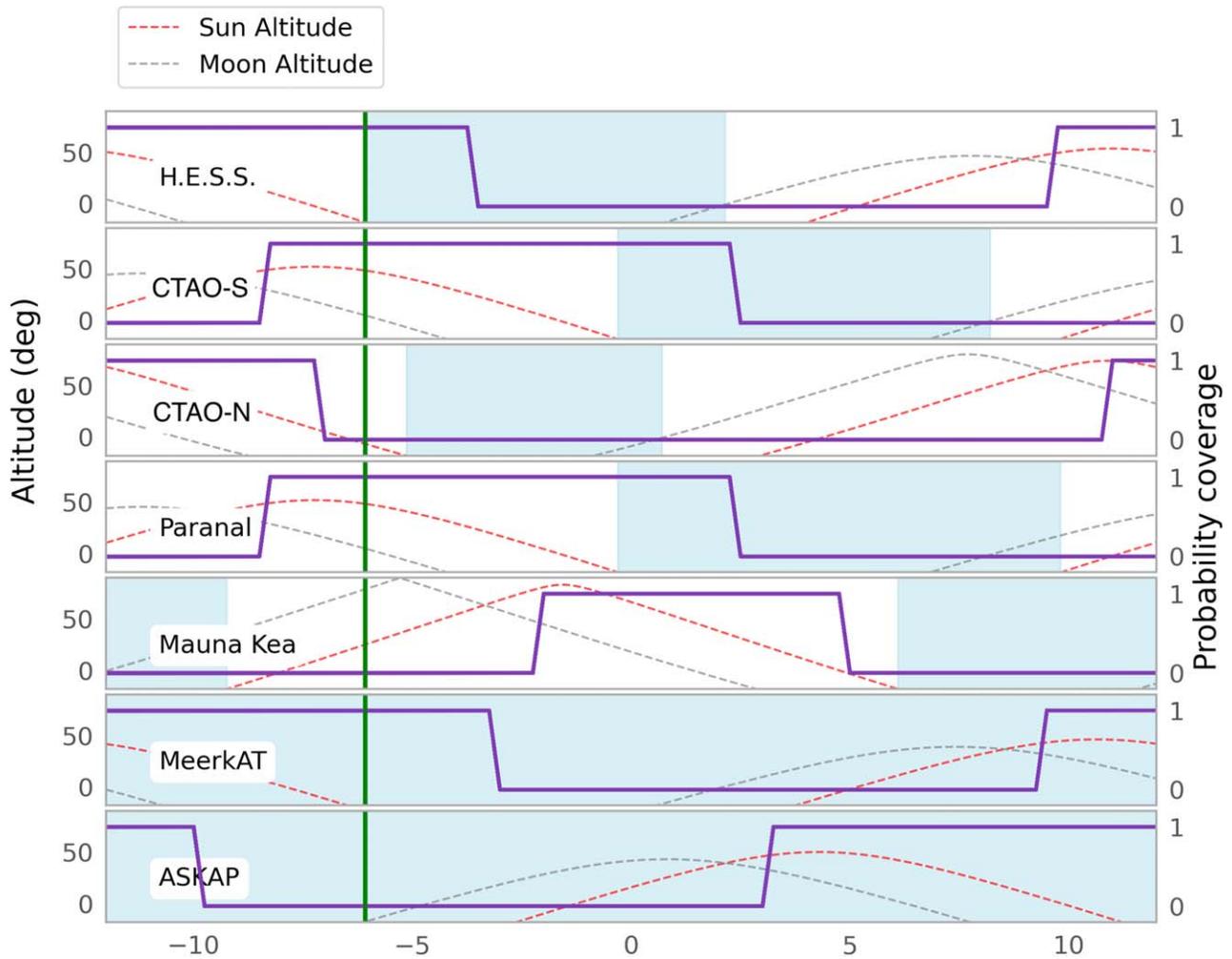

**Figure 1.** Observability and visibility considerations for GW170817 on the night of 2017 August 17–18 for seven observation sites: H.E.S.S., CTAO South, CTAO North, Paranal, Maunakea, MeerKAT, and ASKAP. The localization region is an update computed with the LALInference pipeline (Veitch et al. 2015). The dotted lines show the altitudes of the Sun and Moon as a function of time in each site respectively. The light blue areas represent the times when observation is possible for each observatory (computed with 15 minute time bins). The green lines show the time of the distribution of the alert. From the bottom, radio telescopes are considered to operate during day, dark, and Moon time. Optical telescopes are considered to operate during dark and Moon time, while IACTs are here considered to operate only during darktime. Darktime is defined by the Sun (Moon) being at least 18° (0°5) below the horizon. The violet line shows the fraction of the probability map that is accessible to the site at a given time.

### 2.4. N-observatory/N-telescope Observation Scheduling

`tilepy` enables deriving observation plans for multiple-observatory and multiple-telescope follow-ups. The code allows the combination of multiple telescopes on the same site or different sites regardless of the temporal and spatial simultaneity of the observations. For each telescope or observatory site, a configuration file with the specific telescope parameters and visibility constraints must be provided. The only common feature is the astrophysical event and the start time of the observation campaign. The observation scheduling process is straightforward and follows a *greedy scheduling* approach: the initial observatory that is capable of observing right away takes priority. Subsequently, as other observatories become available for scheduling observations, the masked sky map that excludes the already scheduled observations is utilized to plan their respective observations. The objective is to optimize the probability coverage in the shortest possible timeframe. Consequently, a series of observations will be scheduled for each telescope, considering factors such as its specific observability and visibility conditions, operational characteristics like FoV and zenith angle preferences, and the desired depth/duration of the observations. This methodology applies to N-telescopes and N-observatories in a generalized manner.

### 2.5. Additional Features

`tilepy` incorporates supplementary features that augment the core functionalities of the scheduling code. These enhancements allow for a more customized adaptation of the observation planning to address specific requirements of various scientific use-cases. The details of these features are outlined below:

- *Altitude optimization.* `tilepy` provides the option to use a weighting of the zenith angle of the various proposed pointings. Typically, this is used to favor low zenith angle, i.e., low atmospheric observations over high zenith angle ones. The weights are employed to assess the covered probability in two consecutive zenith angle layers, in steps of 5°. If the weighted probability is greater in the layer with the larger zenith angle, it is selected over the one in the subsequent zenith angle step.

*Deep exposures.* The regions with the highest probability are prioritized for coverage initially if we disregard the observability and visibility constraints. As the observing campaign progresses, the probability density that remains to be covered can drop significantly, particularly for events with a steep probability density profile. `tilepy` includes an option that would stop the initial observation plan when a minimum achievable coverage per pointing is not reached any longer. The observation plan will then include revisits of high-probability regions, even if they have been previously observed. This feature allows to obtain additional observations of the most interesting regions, which is an option to obtain data under better observation conditions (e.g., low zenith angles, less moonlight, etc.) or even scientifically motivated cases as that of searches for late afterglow emission in GRBs.

*Minimum probability coverage.* `tilepy` offers the flexibility to schedule a pointing only when a user-defined probability coverage is attained by that observation. This feature helps to avoid spending valuable telescope time on low-probability regions.

*Observation ranking.* The ordering of the various pointings of an observation campaign provided by `tilepy` might not be respected by observers, owing to factors such as inclement weather or instrumental failures that could lead to missing the optimal observation time for each pointing. To account for this, the user is provided with a secondary schedule wherein observations are arranged based on the probability coverage of individual pointings. Each pointing is assigned an observational priority. The probability of a region that is covered more than once is counted independently of the coverage of any overlapping pointings. This contrasts with the primary schedule, where already-covered regions are disregarded to avoid overestimating the probability coverage of a pointing through double-counting. Furthermore, each pointing in the secondary schedule is accompanied by an observational window, tailored to the observatory's location, observational constraints, and the evolving accessible sky window throughout the night. Observers or burst advocates have thus all the information at hand to flexibly adapt the observation plan without a full recalculation.

*Exclusion of previously observed pointings.* In some cases, like for Fermi-GBM and GW alerts, initial transient event alerts can be superseded by updated information during the ongoing follow-up campaign. `tilepy` provides the option of masking the regions that were already observed and thus excluding them from a renewed probability computation. This avoids covering the same region multiple times.

*Visualization plots.* The user can enable visualization tools embedded into the `tilepy` code. The user will be provided, in addition to the scheduling, with figures of the evolution of the telescope visibility with time, the evolution of the altitude of the scheduled observations throughout the night, the evolution of the zenith angle of the pointings with time, and summary plots of the proposed pointings. In the case of planning a campaign involving multiple observatories, these tools are provided for each telescope and for the combined observational strategy.

*Multiduration exposures.* In the general case, the user-defined duration of each observation is constant for a given telescope. An additional feature allows the user to define a set of observation durations for the observation campaign, which will be used in the scheduling of each pointing. This is useful to follow-up transient sources that may display fading-like behavior, allowing the adjustment of observation windows based on the anticipated evolution of the source. In this case, the user would possibly request later observations to have a longer duration and thus ensure a constant detection likelihood.

## 3. Architecture and How to Use

The `tilepy` is a `python`-based code that follows a modular structure that makes it easy to understand, maintain, and update. A scheme of the code is shown in Figure 2, where the main features explained in Section 2 are depicted. The input information to run `tilepy` consists in a `HEALPix` sky map (either via a URL or a local `fits` file), the starting time of the observation campaign, and observation configuration files. A separate configuration file is necessary for each participating telescope. It includes the observatory location, visibility and observability parameters, as well as parameters related to the sky map treatment, connected to the features mentioned in the previous section. These include the choice of algorithm, the high- and low-resolution `NSIDE`, zenith angle weighting, minimum values for the 2D or 3D probabilities (used to give priority to sky locations), the percentage of the sky localization to be considered in the scheduling, as well as parameters connected to the use of a galaxy catalog, as the maximum event distance up to which one would use it and the use of the approach outlined in Ducoin et al. (2020). In this way, the user can set strategic preferences in the configuration file through activating the available flags. The core functions are organized in three levels, from high to low: the observation scheduler, the tiling determination, and the pointing tools. The user script calls the top-level functions in the observation scheduler, which parse the main parameters and assesses the best strategy to use according to the inputs of the user and the configuration file parameters. The various tiling strategies are integrated in the tiling determination stage. These functions are in charge to obtain the full observation scheduling, including times, coordinates of the pointings, and the covered probability. The low-level functions needed for this goal are called pointing tools.

The `tilepy` code, an installation guide, and how-to-use guide can be found on https://github.com/astro-transients/tilepy. Examples on how to produce complex observation campaigns for a variety of science cases, including GW, GRBs, and neutrino, are covered via Jupyter notebooks. A set of standalone useful tools for multimessenger and time-domain astronomy is also provided at the user's convenience, also via Jupyter notebooks and scripts. These focus on the visualization of observation campaigns and summary plots. Scripts to reduce and convert the galaxy catalog to `HD5F` format and visualization tools of a hypothetical source and sky maps, and Moontime and darktime computation and accessible sky visualization are also included. `tilepy` can also directly be run through its application programming interface (API) found on https://tilepy.com. On the same website, detailed code and API documentation are provided. Finally, `tilepy` functionalities are also integrated into the Astro-COLIBRI platform[5] and can thus be used directly from its front-end interface (Reichherzer et al. 2021). A dedicated help desk and discussion forum are available with the Astro-COLIBRI forum at https://forum.astro-colibri.science/c/instrumentation-and-tools/tilepy.

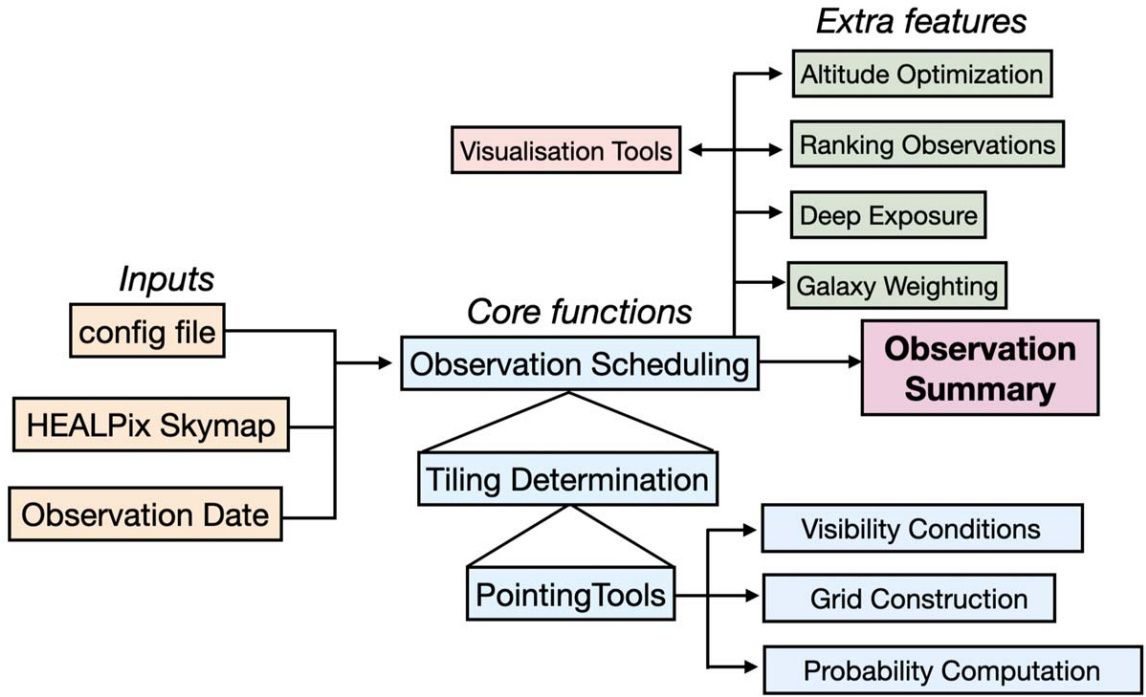

**Figure 2.** Overview of the main workflows within the `tilepy` Python package.

**Table 1**
Summary of the Telescope Configuration Used for the Follow-up Campaign of the GRB 231012A Trigger

| Telescope | Site | Strategy | FoV$_{radius}$ (deg) | $\theta_{max}$ (deg) | $P_{max}$ | $P_{duration}$ (min) | Observation Cond. |
|---|---|---|---|---|---|---|---|
| LST1 | LP | 2D integrated | 2.5 | 70 | 20 | 15 | Dark and moderate Moon |
| LST2 | LP | 2D integrated | 2.5 | 70 | 20 | 15 | Dark and moderate Moon |
| LST3 | LP | 2D integrated | 2.5 | 70 | 20 | 15 | Dark and moderate Moon |
| LST4 | LP | 2D integrated | 2.5 | 70 | 20 | 15 | Dark and moderate Moon |

### 4. Usage Examples

#### 4.1. Very Poorly Localized Fermi-GBM GRB: Multitelescope Campaign at One Site

We illustrate the case of an array of mid-FoV telescopes following a very poorly localized GRB using the Fermi-GBM detection of GRB 231012A (Fermi GBM Team 2023). We consider a site at the Roque de los Muchachos Observatory (ORM) on La Palma (Spain) with an array of IACTs divided into four subarrays: LST1, LST2, LST3, and LST4. We schedule observations independently for the four subarrays, allowing for dark and moderate Moon-time observations. All four subarrays have the same configuration as described in Table 1. The strategy consists of following the 2D probability distribution and sorting the values obtained according to the integrated probabilities within the FoV. The start time of the observation campaign is set to 2023 October 12 19:42 UTC. As the campaign starts, all telescopes start to observe in parallel and cover the localization region in a synchronized way. Each pointing with a fixed duration of 15 minutes is optimized for maximum probability coverage at the earliest possible observation time. Using the `tilepy` observation plan, we obtain the observing strategy presented in Figure 3. A total 80% coverage of the sky map can be covered in 1.5 hr, as shown in Figure 4. There is an anticorrelation between a larger coverage of a telescope and its telescope ID number, which comes from the followed arbitrary ordering in the scheduling. The improvement of the time required to cover the uncertainty region increases with the number of subarrays $N$; we cover the region 4 times faster than what could be done with all four telescopes observing the same sky portion. The observations last until the minimum coverage required per pointing, set to 1%, is no longer achievable. We note that the division of an array into multiple subarrays presents a trade-off of sensitivity and depth into fast coverage. In the case of the CTAO Northern site, one would probably advocate for stereoscopic observations, i.e., subarrays with a minimum of two telescopes, as well as a combination of LSTs and Medium-Sized Telescopes, the largest telescopes of the CTAO North design.

---
[5] https://astro-colibri.science

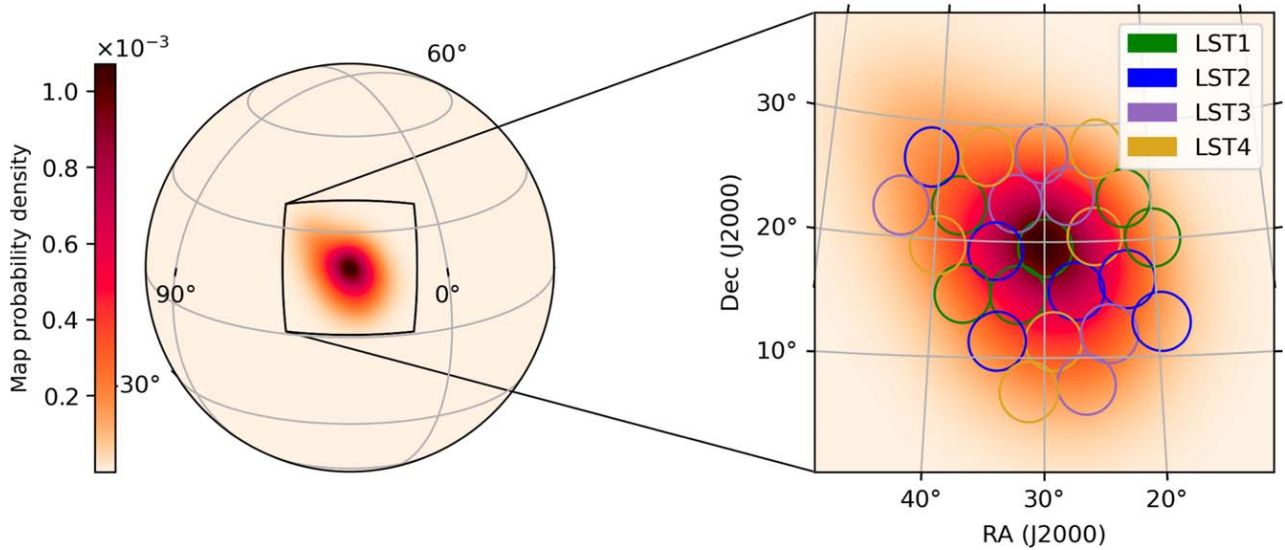

**Figure 3.** Coverage of GRB 231012A trigger with an integrated 2D strategy from four telescopes on the same site, LS1, LST2, LST3, and LST4 in La Palma (LP, see Table 1). The probability map (`NSIDE = 128`) is shown in the globe and inset view. The color bar on the left represents the localization probability in the map. The circles in the inset view represent the telescope pointings. The start of the observation campaign is set at 2023 October 12 17:42:18 UTC.

**Table 2**
Summary of the Telescope Configuration Used for the Follow-up Campaign of the MS230826n Mock GW Event

| Telescope | Site | Strategy | FoV$_{radius}$ (deg) | $\theta_{max}$ (deg) | $P_{max}$ | $P_{duration}$ (min) | Observation Cond. |
|---|---|---|---|---|---|---|---|
| CTAO-N | LP | 3D integrated | 2.0 | 70 | 20 | 15 | Dark and moderate Moon |
| CTAO-S | Paranal | 3D integrated | 4.0 | 60 | 20 | 10 | Dark and moderate Moon |

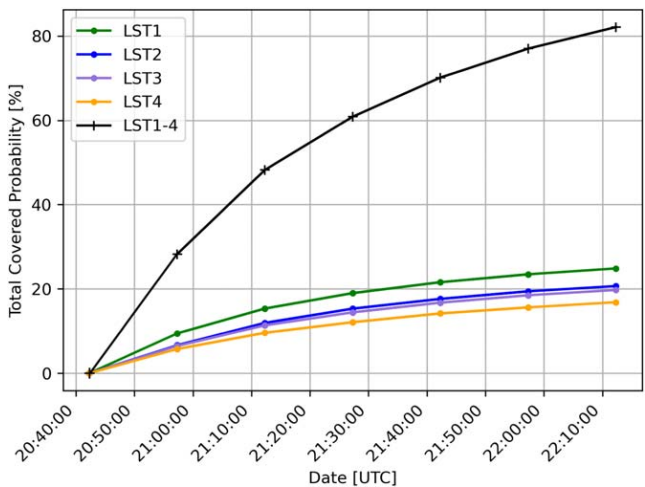

**Figure 4.** Comparison of the cumulative probability covered in the GRB 231012A trigger follow-up campaign by the telescopes independently and combined.

### 4.2. GW Follow-up: Multiobservatory Campaign from the Northern and Southern Hemisphere Using the FoV-integrated 3D Strategy

We schedule observations on a poorly localized simulated BNS GW event (event 927563 from Singer et al. 2016) at a distance of $89.66 \pm 19.29$ Mpc. In this case, two gamma-ray observatories are selected, one on each hemisphere: CTAO North at ORM, La Palma (Spain), and CTAO South, at Paranal Observatory (Chile). Each site has thus a different accessible sky at a given time. The telescope configuration is summarized in Table 2. We allow for dark and moderate Moon-time observations at both sites. The GW event is close enough and contains luminosity distance information allowing the selection of a 3D, FoV-integrated probability strategy. A minimum probability coverage cut at 0.5% and 2% for CTAO-N and CTAO-S respectively is set for each pointing to be scheduled. We chose these values since the assumed FoV of CTAO-N is smaller than CTAO-S leading to smaller integrated probabilities. The observation campaign starts at 2023 March 15 10:30:10. Both sites cover the localization regions that are reachable during their respective observing hours. A representation of the achieved coverage is shown in Figure 5. The complexity of the campaign is illustrated as observation conditions for CTAO-N are met before CTAO-S. We see that, hours after the start of the observational campaign, CTAO-N starts covering the most probable parts of the uncertainty region, then CTAO-S fills the outer gaps. Both observatories stop observing when the follow-up observation conditions are no longer met. The joint observations by CTAO North and CTAO South cover 86% (as shown in Figure 6) of the galaxy probability with a total of 20 pointings. The regions with the largest concentration of probable galaxies to host the GW event are successfully covered.

### 4.3. GW Follow-up: Multioptical Observatory Campaign around the Globe

For the case shown in Figure 7, we consider the GW190814 (Abbott et al. 2020) localization map. The GW event is well localized to 18.5 deg$^2$ at 90% confidence level.

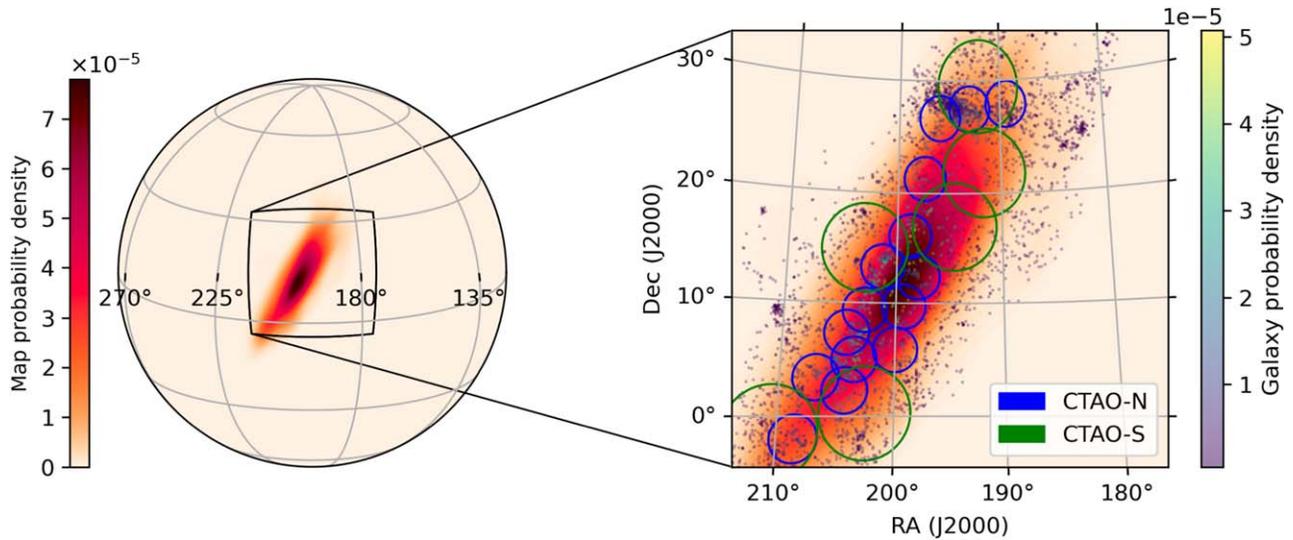

**Figure 5.** Coverage of the GW simulated map (event 927563) taken from Singer et al. (2016) with an integrated 3D strategy from two sites, CTAO North and CTAO South (see Table 2). The probability map (NSIDE = 512) is shown in the global and inset view. The color bar on the left represents the localization probability in the map. The color bar on the right represents the galaxy probability after correlation with the probability map. The galaxies having a probability of more than 1% the maximum galactic probability are represented in the inset view by dots. The circles in the inset view represent the telescope pointings. The start of the observation campaign is set at 2023 March 15 10:30:10 UTC.

We consider seven hypothetical observatories and telescopes on six different sites: two in Paranal (ESO and ESO2), two in La Palma (LP and LP2), one in the Observatoire de Haute-Provence in France (OHP), one in South Africa (SA), and one in Hawaii (HA) with stringent constrains on the allowed number of observations. The diverse telescope configurations are summarized in Table 3. The scheduling is set to start at 2023 September 15 01:30:10 UTC. The last telescope to join the campaign as the observability and visibility conditions are met on its site is the one in Maunakea (Hawaii). The total coverage with all telescopes is 93% (as shown in Figure 8) of the total galaxy probability with a total telescope observation time varying from 30 to 90 minutes.

### 4.4. IPN GRB Follow-up: Single Narrow FoV Radio Telescope in the Southern Hemisphere

We consider the high-resolution localization region of GRB 20120612[6] provided by IPN (Hurley et al. 2010) to be covered by the ATCA radio telescope at the Paul Wild Observatory in Australia. We assume a telescope system with a 0.05° FoV, a maximum zenith angle observation of 60°, 10 minute duration per pointing, and a total of 50 allowed pointings. The observation campaign is set to start at 2017 August 17 10:30:10 UTC. A total coverage of 71% is achieved with 50 pointings where a portion of them is scheduled for daytime observations. The pointings are shown in Figure 9. We find that, with only 20 pointings, we cover already more than 40% of the localization map. In reality, the observers can choose to limit the number of requested pointings or set a minimum probability coverage per pointing.

### 5. Conclusion and Outlook

In conclusion, tilepy is a novel tool for the scheduling of follow-up observations of poorly localized astrophysical (transient) events. It is designed to provide users with an optimized observation plan and offers maximum flexibility. It is fine-tuned for automatic follow-ups but is also suited for manual usage. After the initial configuration is provided by the user, tilepy has the capabilities of being able to independently make decisions on the follow-up strategies without any further human intervention. With an easy-to-use telescope configuration, a start date, and a link to the astrophysical event localization, tilepy will assess telescope observability and visibility requirements, select the most suited strategy, schedule the pointing pattern in function of time maximizing the covered probability, and provide the user with schedules and visual aids. The architecture of tilepy allows it also to be used in a deeply customized way, allowing the user to define their preferences, either through parameter inputs, additional features, or specific changes to the code. The functionalities of tilepy make it an ideal scheduling tool for poorly localized events that can be integrated into various automatic transient handlers with the least-possible effort. tilepy is currently used in the H.E.S.S. (Hoischen et al. 2022; Ashkar et al. 2021) and CTAO-LST (Carosi et al. 2021) collaborations. tilepy has its own API, hosted at https://tilepy.com. tilepy has been embedded in Astro-COLIBRI, accessible from the browser at https://astro-colibri.com or directly from the app, readily available on current app download platforms.

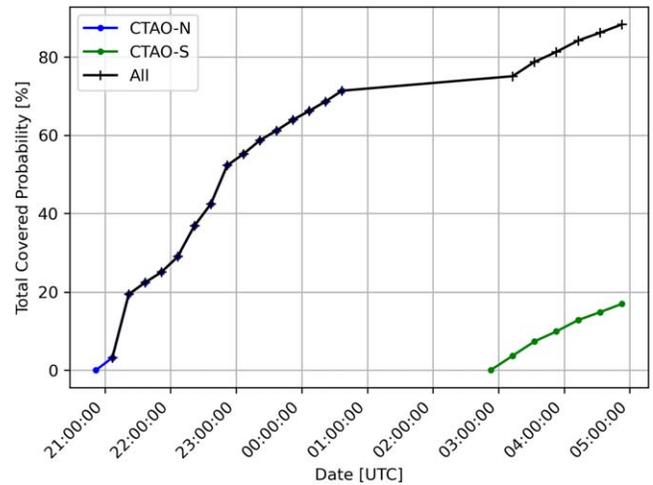

**Figure 6.** Comparison of the cumulative galaxy probability covered in the GW simulated event 927563 follow-up campaign by the telescopes independently and combined.

---

[6] https://www.ioffe.ru/LEA/ShortGRBs_IPN/data/20120612T59382/GRB20120612_T59382_IPN_map_hpx.fits.gz

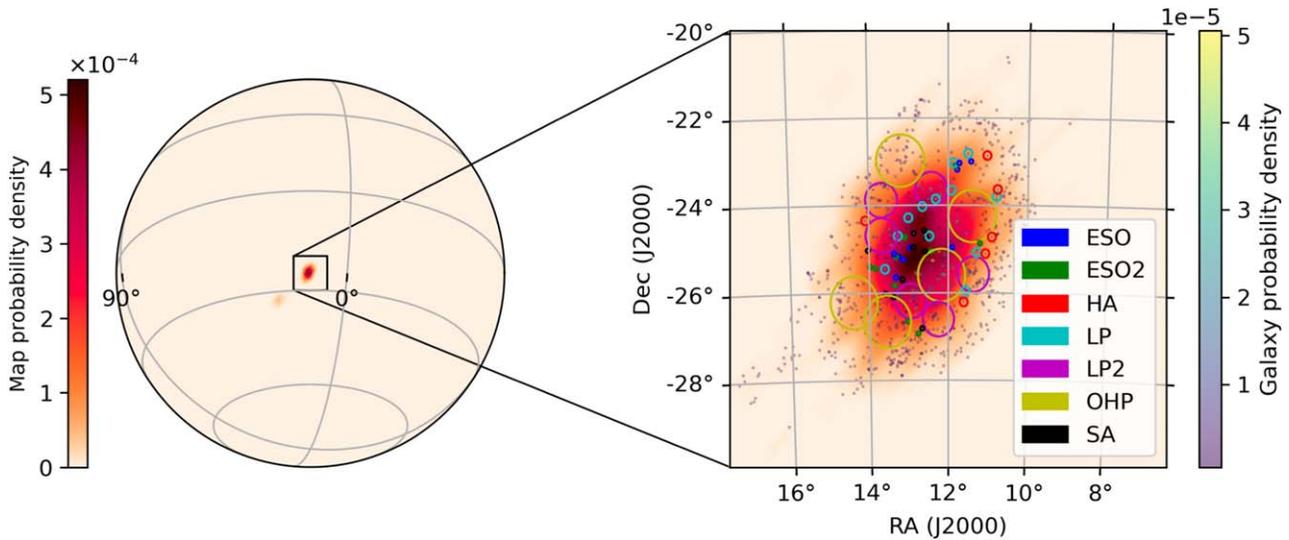

**Figure 7.** Like Figure 5, but showing coverage of GW190814bv (NSIDE = 1024) with an integrated and targeted 3D strategy from seven telescopes on five different sites for which observations could be scheduled (see Table 3). The start of the observation campaign is set at 2023 September 15 01:30:10 UTC.

Table 3
Summary of the Telescope Configuration Used for the Follow-up Campaign of the GW190814bv Event Trigger

| Telescope | Site | Strategy | FoV$_{radius}$ (deg) | $\theta_{max}$ (deg) | $P_{max}$ | $P_{duration}$ (min) | Observation Cond. |
|---|---|---|---|---|---|---|---|
| ESO | Paranal | 3D targeted | 0.05 | 80 | 10 | 5 | Dark and Moon |
| ESO2 | Paranal | 3D targeted | 0.05 | 80 | 10 | 5 | Dark and Moon |
| LP | LP | 3D targeted | 0.1 | 80 | 12 | 5 | Dark only |
| LP2 | LP | 3D integrated | 0.4 | 70 | 6 | 5 | Dark and Moon |
| OHP | Maunakea | 3D targeted | 0.1 | 80 | 6 | 15 | Dark and Moon |
| SA | South Africa | 3D integrated | 0.6 | 80 | 6 | 5 | Dark and Moon |
| HA | Hawaii | 3D targeted | 0.05 | 80 | 10 | 5 | Dark and Moon |

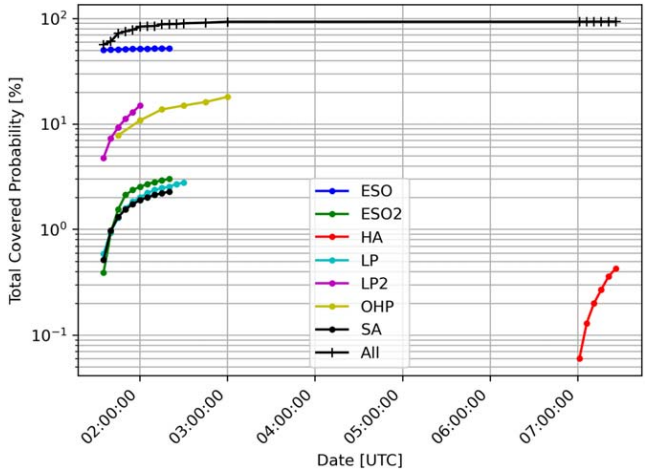

**Figure 8.** Comparison of the cumulative galaxy probability covered in the GW190814bv trigger follow-up campaign by the telescopes independently and combined.

tilepy is evolving as the multimessenger field broadens, and new instruments are being built. Due to the instrument response function of GW detectors, the next generation of GW detectors, led by Einstein Telescope and Cosmic Explorer, will still have limited capabilities in source localization, with a large number of events having localization uncertainties beyond hundreds of squared degrees (Ronchini et al. 2022). Even more interesting will be the follow-up of premerger GW alerts, as a way to drive electromagnetic counterpart searches before the merger happens (Banerjee et al 2023). Therefore, tiling strategies are poised to endure for an extended period, particularly in BNS follow-up observations. Regarding the follow-up of further messengers, there is the neutrino case, already supported in tilepy. Further boost in neutrino follow-up observations could be obtained by considering catalogs of the most likely sources, as done in GW follow-ups. For this purpose, catalogs from AGNs could be used. Further updates of the algorithm that are being considered include the implementation of increased flexibility of exposure times for individual pointings, possibly computed based on the assumption of the energy spectrum and lightcurve of the transient phenomena to be observed. tilepy is customized to tackle this case, to study the optimal exposure to detect the source in GW follow-ups by the next generation IACT, the Cherenkov Telescope Array Observatory (Green et al. 2023). A generalization of this custom method to generic spectrum, lightcurve, and instrument response function, to obtain the exposure per pointing required for the source detection, is planned for the near future.

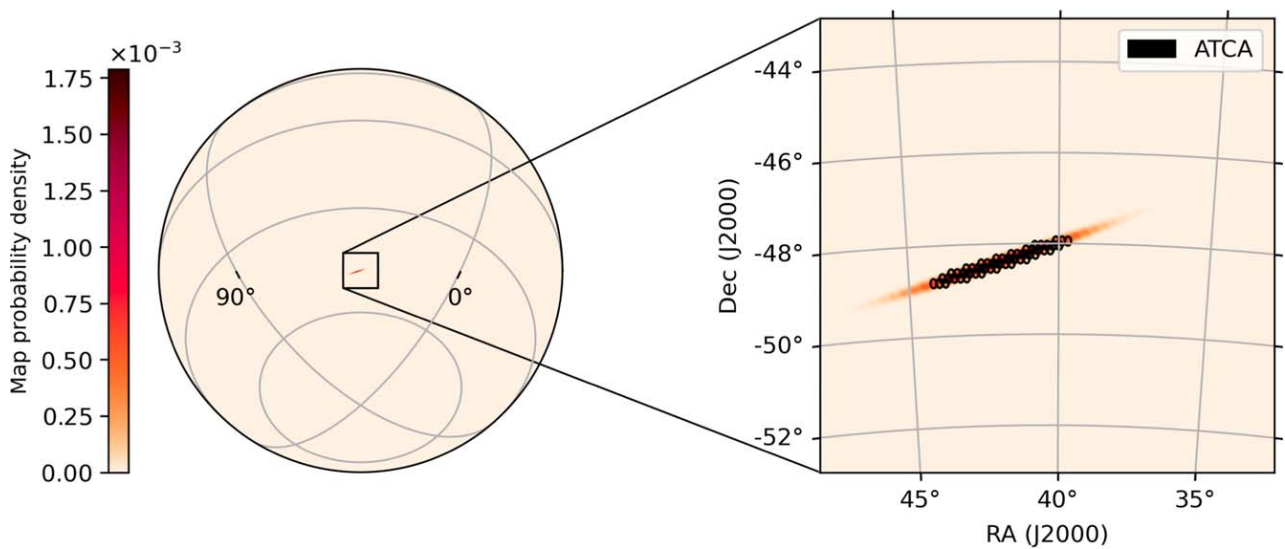

**Figure 9.** Coverage of IPN localization map of GRB 20120612 (`NSIDE` = 2048) with a radio observatory at the Paul Wild Observatory in Australia. The probability map is shown in the globe and inset view. The color bar on the left represents the localization probability in the map. The circles in the inset view represent the telescope pointings. The start of the observation campaign is set at 2017 August 17 10:30:10 UTC. Note that this time has been selected for illustrative purposes, as it corresponds to a period for which observation windows are found at the Paul Wild Observatory.

Another major update will be the handling of different shapes of telescope FoV, as the current implementation focuses in the case of circular FoV. A special handling of equatorial mount telescopes with noncircular FoVs is ongoing. Future plans include the integration of different optimization algorithms at the end of the scheduling, following an approach similar to that of the traveling salesman problem. Some of the methods described in this paper targeted the reduction of computation time, and are efficient strategies that show improvements of orders of magnitude, as is the case when rebinning high-resolution sky maps. Still, the complexity of large multiobservatory campaigns increases the computation time. One key improvement includes the further reduction of the computation time in complex observation campaigns. Other major improvements encompass the management of space observatory scheduling as well as the handling of alternative forms of localizations beyond `HEALPix`.

`tilepy` is continuously undergoing upgrades and is being actively adapted to the changes in the multimessenger domain. These are accessible both in the tilepy GitHub and in the Zenodo repository (Seglar-Arroyo et al. 2024). We invite feedback and requests from the community and encourage discussions in our dedicated help desk and discussion forum that is available with the Astro-COLIBRI forum at https://forum.astro-colibri.science/c/instrumentation-and-tools/tilepy.

### Acknowledgments

The `tilepy` package relies on the following Python packages: astropy (https://github.com/astropy/astropy), scipy (https://github.com/scipy/scipy), healpy (Górski et al. 2005; Zonca et al. 2019), matplotlib (https://github.com/matplotlib/matplotlib), MOCpy (https://github.com/cds-astro/cds-moc-rust), numpy (https://github.com/numpy/numpy), pandas (https://github.com/pandas-dev/pandas), pytz (https://github.com/stub42/pytz), ephem (https://github.com/brandon-rhodes/pyephem), gdpyc (https://github.com/ruizca/gdpyc), fastparquet (https://github.com/dask/fastparquet), skyfield (https://github.com/skyfielders/python-skyfield), and ligo.skymap (https://github.com/lpsinger/ligo.skymap).

We thank the Astro-COLIBRI team for hosting and maintaining the `tilepy` API. We thank Enrique Garcia for useful insights, which were pivotal in transforming the code into a `python` package. We thank Nicolas Leroy and Luis Fariña for their insightful comments, which enhanced the quality of this paper. M. S.A. is supported by the grant FJC2020-044895-I funded by MCIN/AEI/10.13039/501100011033 and by the European Union NextGenerationEU/PRTR. We acknowledge support by Institut Pascal at Université Paris-Saclay during the 2nd Astro-COLIBRI Multimessenger Astrophysics workshop 2023, with the support from the Unistellar citizen science program, IRFU CEA Paris-Saclay, and AS OV. We also acknowledge the Paris-Saclay Astroparticle Symposium 2023, with the support of the P2IO Laboratory of Excellence (program "Investissements d'avenir" ANR-11-IDEX-0003-01 Paris-Saclay and ANR-10-LABX-0038), the P2I axis of the Graduate School of Physics of Université Paris-Saclay, as well as IJCLab, CEA, IAS, OSUPS, and the IN2P3 master projet UCMN. We furthermore acknowledge the support of the French Agence Nationale de la Recherche (ANR) under reference ANR-22-CE31-0012 and by the Programme National des Hautes Energies of CNRS/INSU with INP and IN2P3, cofunded by CEA and CNES.

### ORCID iDs

Monica Seglar-Arroyo ● https://orcid.org/0000-0001-8654-409X
Halim Ashkar ● https://orcid.org/0000-0002-2153-1818
Mathieu de Bony de Lavergne ● https://orcid.org/0000-0002-4650-1666
Fabian Schüssler ● https://orcid.org/0000-0003-1500-6571